\documentclass{llncs}

\usepackage{amsmath,amssymb,latexsym}
\usepackage{graphicx}
\usepackage{epstopdf}
\usepackage{algorithm}
\usepackage{algorithmic}
\usepackage[numbers,sort&compress]{natbib}
\usepackage{natbib}
\usepackage{multicol}
\usepackage{array}
\usepackage{multirow}
\usepackage{subfig}

\newcommand{\ro}{\mathcal{R}}
\newcommand{\enm}{\alpha} 
\newcommand{\lnm}{\beta}  
\newcommand{\dmc}{\mathcal{M}}
\newcommand{\pr}{\mathbb{P}}

\newcommand{\fav}{\succ}
\newcommand{\hist}{\mathcal{H}}

\newcommand{\real}{\rm \tiny real}
\newcommand{\mb}{\rm \tiny MLN}
\newcommand{\cg}{\rm \tiny CG}
\newcommand{\cgr}{\rm \tiny CGR}
\newcommand{\rand}{\rm \tiny rand}
\newcommand{\tseq}{\gamma}

\makeatletter
\renewcommand\bibsection%
{
  \section*{\refname
    \@mkboth{\MakeUppercase{\refname}}{\MakeUppercase{\refname}}}
}
\makeatother

\begin{document}

\mainmatter  

\title{Reconstruction of Network Evolutionary History from Extant Network Topology and Duplication History}

\titlerunning{Reconstruction of Network Evolutionary History}

\author{Si LI\inst{1}, Kwok Pui CHOI\inst{1,2}, Taoyang WU\inst{1}, Louxin ZHANG\inst{1}}

\authorrunning{S. LI et al.}

\institute{Department of Mathematics,
\and
Department of Statistics and Applied Probability,\\
 National University of Singapore,
 Singapore 119076\\
 \email{\{g0800874,stackp,matwt,matzlx\}@nus.edu.sg}}

\toctitle{}
\tocauthor{}
\maketitle
\graphicspath{{./}}

\begin{abstract}

Genome-wide protein-protein interaction (PPI) data are readily available thanks to recent breakthroughs in biotechnology. However, PPI networks of extant organisms are only snapshots of the network evolution. How to infer the whole evolution history becomes a challenging problem in computational biology. In this paper, we present a likelihood-based approach to inferring network evolution history {  from} the topology of PPI networks and the duplication relationship among the paralogs.  
Simulations show that our approach outperforms the existing ones in terms of the accuracy of reconstruction. 
Moreover, the growth parameters of several real PPI networks estimated by our {  method} are more consistent with the ones predicted in literature. \\

\end{abstract}

\section{Introduction}

Recent progress in experimental systems biology provides us with an unprecedented amount of genome-wide protein-protein interaction (PPI) data~\cite{hakes08}. In order to obtain a deeper insight into the molecular machinery behind these interactions, many network models have been proposed to study or model PPI evolution~\cite{bara04,yamada09,stumpf07}. However, PPI networks of extant organisms are only snapshots of network evolution, and inferring the whole network evolution history remains a challenging problem in computational biology~\cite{Navlakha11}.

Unlike many networks studied in technology and sociology, the main growth mechanism of PPI network is gene duplication and divergence~\cite{wagner01}:   when a new node is added to the network, it copies all the interactions of an existing node designed as the anchor node;  subsequently some edges adjacent to {  one of} these two nodes are randomly lost.  This mechanism was explicitly converted to a network growth model by Vazquez et al. in~\cite{vazquez01}.
 Since then many extensions have been put forth, see for examples, \cite{chung03,
 sole02,pastor03,gbebek06,
 bahan02}. Here we shall focus on a particular one called duplication-mutation with complementarity (DMC), which is the best model to fit the {\em D. melanogaster} (fruit fly) PPI network according to a recent study by Middendorf et al.~\cite{manuel05}.

 When a growth model is fixed, the problem of reconstructing the evolutionary history of an observed network is to infer the relative order of the nodes according to which the network evolved (see Section~\ref{sec:hist} for definitions). Better understanding of this problem {  can provide further insights into not only how PPI networks are formed, but also how they will possibly evolve in the future}. Several approaches to {  address} this problem have been proposed in recent years.  In order to obtain better ways of predicting protein modules, Dutkowski and Tiuryn introduced a Bayesian network framework to infer the posterior probability of interactions between ancestral nodes based on a duplication and speciation model~\cite{dutko07}. A similar approach was used by Pinney~\cite{pinney07} to infer ancestral interactions between bZIP proteins. Gibson and Goldberg  proposed a merging algorithm to reconstruct the evolutionary history of PPI networks using gene trees~\cite{gibson09}. A novel framework for estimating the topology of the ancestral networks based on maximal likelihood
{  was} presented by Navlakha and Kingsford in~\cite{Navlakha11}. {  Recently,} Patro et al. \cite{Rob2011} used a maximal parsimony approach that appends edges in observed networks to duplication history forest.

Here we introduce a new history inferring framework based on the maximal likelihood principle.  In contrast to {  the model-based methods in}~\cite{Navlakha11}, our approach incorporates not only the topology of observed networks, but also the duplication history of the proteins contained in the networks. Although the evolution of topology is often
determined by some growth mechanisms, the duplication history of the proteins can be inferred independently by phylogenetic studies~\cite{pinney07,Rob2011}.
After establishing some theoretical results concerning the DMC model, we reduce the problem of finding most probable history of ancient networks to an optimization problem, and propose some efficient heuristic algorithms to solve the latter problem.
Simulations show that our method provides better inference than the ones in~~\cite{Navlakha11}. Moreover,   we also applied our algorithm to the PPI networks of {\em S. cerevisiae} (budding yeast), {\em D. melanogaster} and {\em C. elegans} (worm), and the growth parameters obtained by our approach are {  more} consistent with the ones predicted in~\cite{wagner01,nadia06}. 
 Finally, we also propose an improved measure for comparing two histories.

The rest of the paper is organized as follows: Section \ref{sec:method} provides the framework of reconstruction, including the technical background and the inference method.
 In Section \ref{sec:result} we present the inference results for simulations and real data sets. We conclude in Section \ref{sec:discuss} with a brief discussion and some possible related research directions.


\section{Methods}\label{sec:method}

\subsection{Modeling Network Evolution}
\begin{figure}
    \begin{center}
        {\resizebox{0.9\columnwidth}{!}{\includegraphics{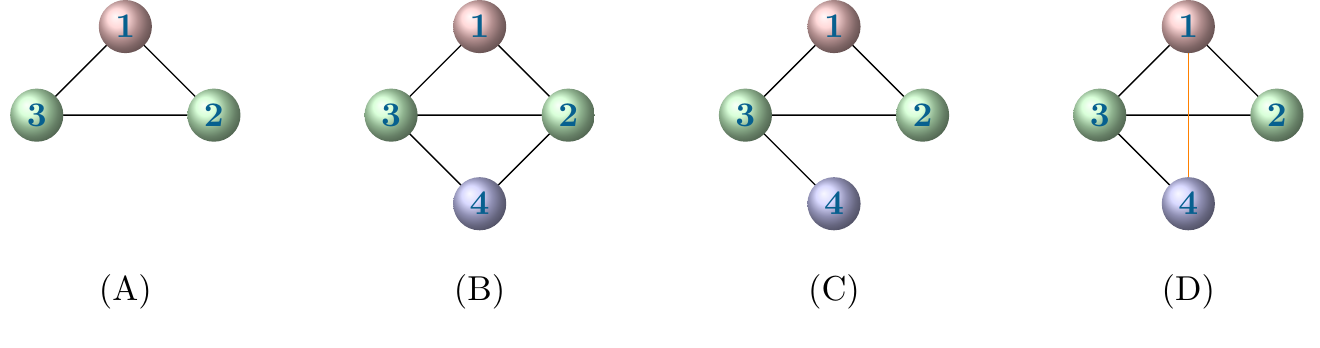}}}
        \caption{
             Illustration of the DMC model. (B) is obtained from (A) by one duplication step, with node 1 (represented in maroon) as the anchor node and node 4 as the duplicate node (represented in purple); the probability that node 1 is chosen as the anchor node is 1/3 because the network in (A) contains three nodes. (C) is obtained from (B) by the mutation step, which occurs with probability $p(1-p)/2$. (D) is obtained from (C) by the complementarity step,which occurs with probability $p_c$.}
           \label{fig:dmc}
           \end{center}
\end{figure}

In the DMC model $\dmc:=\dmc(p_c,p)$, where
$p_c$ and $p$ are the two parameters that specify the model,
 we start with an initial graph $G_0$, the so-called seed graph.  At each time step $t$, the graph $G_t$ is obtained from $G_{t-1}$ by the following procedures (see Fig.~\ref{fig:dmc} for an illustration):
(1) (Duplication) A node $u_{t}$ is chosen uniformly at random from the set of nodes in $G_{t-1}$, and a new node $v_t$ is added and connected to every neighbor of $u_t$. Here $u_t$ and $v_t$ are often referred to as the anchor node and duplicate node at step $t$, respectively.
(2) (Mutation) For each neighbor of $u_t$, say $w$, we choose one edge from $(u_t,w)$ and $(v_t,w)$ with equal probability, and this chosen edge is deleted with probability $1-p$.
(3) (Complementarity) The nodes $u_t$ and $v_t$ are connected with probability $p_c$.

Note that the DMC model is Markovian, that is, the probability of obtaining $G_t$ when $G_{t-1}$ is given depends {  solely} on the parameters of $\dmc$. For example, denoting the network (A) and (D) in Fig~\ref{fig:dmc} by $G_{t-1}$ and $G_t$, respectively, then the probability $\pr(G_t|G_{t-1},\dmc)$ that $G_t$ is evolved from $G_{t-1}$ by one step under the model $\dmc$  is $p(1-p)p_c/2$.



\subsection{History Reconstruction}
\label{sec:hist}

Given an observed network $G$, a {\em growth history} $\hist$ of $G$ is a graph sequence $(G_0,G_1,\cdots,G_n)$ such that $G_n=G$ and for each index $t$ in $\{1,\cdots,n\}$, the graph $G_t$ {  can be obtained} from $G_{t-1}$ in one step under the DMC model $\dmc$. The first graph $G_0$ is referred to as the seed graph of the history. In addition, the number $n$ is called the {\em span} of the history. Clearly, a history $\hist$ induces a unique sequence $\theta:=\theta(\hist)$ of duplicate nodes, that is, $\theta=(v_1,\cdots,v_n)$ such that for all $t$, node $v_t$ is the {  unique} node in $G_t$, but not $G_{t-1}$.

 Given a network $G$, let $\hist$ be the growth history we hope to infer. The probability of $G$ being evolved according to history $\hist$, when viewed as a function of the unknown history $\hist$, is called the {\em likelihood function} $L(\hist\,|\,G,\dmc)$ that is given by
$$
L(\hist\,|\,G,\dmc)=\prod_{t={1}}^n  \pr(G_t|G_{t-1},\dmc).
$$

We adopt a maximal likelihood approach to infer the history of $G$ as below.

\begin{problem}
Given a network $G$ together with a natural number $n$ and model $\dmc$, construct a growth history $\hist$ that maximizes the likelihood $L(\hist\,|\,G,\dmc)$ among all histories with span $n$.
\end{problem}

This problem is expected to be difficult since the number of possible histories grows exponentially, {  and we are not aware of any results concerning whether this problem is polynomial-time solvable. } Before introducing a variant of the above problem that is more tractable, we present some necessary tools in the following two subsections.

\subsection{Duplication Forest}


{  We begin with duplication history, which is closely related to network history as gene duplication is a major driving force of PPI network evolution~\cite{wagner01}.  The idea of encoding the duplication history by a forest of binary tree was used  in~\cite{Navlakha11,Rob2011}}. Patro et al. ~\cite{Rob2011} incorporated duplication history in a parsimony approach to reconstruct network history.


A growth history $\hist$ of a PPI network induces a unique duplication forest.  Initially, we have a forest $\Gamma_0$ consisting of isolated nodes that are identical to the set of nodes in the seed graph. At each step $t$,  the forest $\Gamma_t$ is obtained from $\Gamma_{t-1}$ by replacing the anchor node $u_t$ with a cherry $\{u_t,v_t\}$ consisting of $u_t$ and the duplicate node $v_t$.
{  Here a {\em cherry} $\{u,v\}$ is referred to a subtree consisting of two leaves $u$ and $v$ and the internal node adjacent to them}.


 The duplication forest of a PPI network can also be inferred independently without using  its growth history. For instance, such a forest can be reconstructed by the phylogenetic relationships between the genes in the network~\cite{pinney07}. This observation is key to our investigation.

\subsection{Backward Operator}
In this subsection, we will introduce a backward operator that is important in our inference framework.

Consider one step in a growth history, that is, a graph $G_t$ obtained from $G_{t-1}$ in one step by using anchor node $u_t$ and duplicate node $v_t$.
 Now we want to define a backward operator $\ro$ such that $G_{t-1}$ can be determined by this operator and the triplet $(G_t,u_t,v_t)$. To this end, let $\ro_{v_t}^{u_t}(G_t)$ be the graph obtained from $G_t$ by merging the two nodes $u_t$ and $v_t$ in $G_t$, that is, (i) for each neighbor $w$ of $v_t$ such that $w\not =u_t$ and $w$ is not adjacent to $u_t$, {  add an edge $(w,u_t)$}; (ii) delete the node $v_t$ and all edges incident to it.

 Similarly, the backward operator can be applied to the duplication forest, that is, $\ro^{u_t}_{v_t}(\Gamma_t)$ is the forest obtained from $\Gamma_t$ by replacing the cherry $\{u_t,v_t\}$ with the leaf $u_t$. Note that this definition is consistent with the above one in the following sense: If $\Gamma_t$ is the duplication forest corresponding to the network $G_t$, then $\ro^{u_t}_{v_t}(\Gamma_t)$ is the duplication forest associated with $\ro^{u_t}_{v_t}(G_t)$. When the anchor node $u_t$ is clear from the context, we also write $\ro_{v_t}$ for $\ro^{u_t}_{v_t}$.


\subsection{Growth History with Known Duplication Forest}

Using the backward operator introduced above, we shall introduce a scheme to represent a growth history with known duplication forest by a node sequence. Throughout this paper, we use the convention that a node sequence consists of distinct nodes, while a node list may contain repeated nodes.

In general,  a node sequence $\theta=(v_1,\cdots,v_n)$ and a duplication forest $\Gamma$ are said to be {\em compatible} if there exists a (necessarily unique) sequence $(\Gamma^{\theta}_1,\cdots,\Gamma^{\theta}_n)$ of forests such that $\Gamma^{\theta}_n=\Gamma$, and $\Gamma^{\theta}_{t-1}=\ro_{v_t}(\Gamma_t)$ holds for each $t\in \{1,\cdots,n\}$. Note that a necessary and sufficient condition for $\theta$ and $\Gamma$ being compatible is that
$v_t$ belongs to a cherry in $\Gamma^{\theta}_t$ for each $t$.
Denoting the sibling of $v_t$ in $\Gamma^{\theta}_t$, that is, the unique leaf in $\Gamma_t$ that forms a cherry with $v_t$, by $u_t$, we say the list $\pi=(u_1,\cdots,u_n)$ is the {\em anchor} list determined by $\Gamma$ and $\theta$.

As mentioned above, a growth history $\hist=(G_0,\cdots,G_n)$ specifies a duplicate sequence $\theta=(v_1,\cdots,v_n)$ and a duplication forest $\Gamma$. Clearly, the sequence $\theta$ and forest $\Gamma$ must be compatible. On the other hand, given a duplication forest $\Gamma$ associated with a network $G$ and a sequence $\theta$ that is compatible with $\Gamma$, then there exists a unique growth history $\hist$ such that $\theta$ is induced from $\hist$.
In other words, when the duplication forest $\Gamma$ is fixed, a growth history $\hist$ is uniquely determined by the duplicate sequence $\theta$ associated with it. In this context, the likelihood function is defined as
$$
L(\theta\,|\,G,\Gamma,\dmc):=\prod_{i={1}}^n  \pr(G^{\theta}_i\,|\,G^{\theta}_{i-1},\Gamma,\dmc),
$$
where $\pr(G^{\theta}_i|G^{\theta}_{i-1},\Gamma,\dmc)$ is the probability that $G^{\theta}_i$ is evolved from $G^{\theta}_{i-1}$ in one step under the DMC model $\dmc$ and using the anchor node $u_t$ specified by $\theta$ and $\Gamma$. Note that in general the probability  $\pr(G^{\theta}_i|G^{\theta}_{i-1},\Gamma,\dmc)$ is different from
$\pr(G^{\theta}_i|G^{\theta}_{i-1},\dmc)$. Indeed, the latter can be regarded as an ``average" of the former over all possible anchor nodes.


Now, the problem of inferring growth history with given duplication forest,  a variant of Problem 1 that will be studied in this paper, can be formally stated as below.

\begin{problem}
Given a network $G$ together with a duplication forest $\Gamma$ and a growth model $\dmc$, construct a duplicate sequence $\theta$
such that
the likelihood $L(\theta\,|\,G, \Gamma,\dmc)$ is maximized.
\end{problem}

In the above problem, the parameters in the DMC model $\dmc$ are specifically mentioned. However, as we shall see later, the parameters of $\dmc$ are not needed for the history inference problem.

\subsection{Theoretical Results}
\label{sec:theory}

Here we present some theoretical results that are crucial to
solve Problem 2.  
Due to space limitations, all proofs are  outlined in the Appendix.

  \begin{lemma}
  \label{lem:seed:graph}
Given a network $G$ with duplication forest $\Gamma$, for any two sequences $\theta_1$  and $\theta_2$ that are compatible with $\Gamma$, the graph $G^{\theta_1}_{0}$ is isomorphic to $G^{\theta_2}_{0}$.
 \end{lemma}

Given a duplicate sequence $\theta=(v_1,v_2,\cdots,v_n)$, we shall associate it with three families of numbers that are crucial to our analysis. For each duplicate node $v_i$ in $\theta$, let $\delta(v_i)$ be the indicator function that takes value 1 if $v_i$ is connected to its anchor node $u_i$, and $0$ otherwise;
 $\enm(v_i)$  the number of the neighbors shared by $v_i$ and $u_i$; and  $\lnm(v_i):=\lnm(v_i,G^{\theta}_i)$  the number of nodes adjacent to $v_i$ or $u_i$ in $G^{\theta}_i$, but not both.
 Note that $2\delta(v_i)+2\enm(v_i)+\lnm(v_i)$ is equal to the sum of the degree of $v_i$ and that of $u_i$ in $G^{\theta}_i$.

The sum $\delta(\theta):=\sum_{i=1}^n \delta(v_i)$ is called the {\em complementarity number} of history $\theta$,
 the sum $\enm(\theta):=\sum_{i=1}^n \enm(v_i)$ is called the {\em extension number} of $\theta$, and $\lnm(\theta):=\sum_{i=1}^n \lnm(v_i)$ is called the {\em loss number} of $\theta$.

{ 
We complete this subsection with the following two key results. The first one
 shows that the complementarity number and extension number are  constants over all compatible duplicate sequences.}

\begin{theorem}
\label{thm:comp:num}
Given a network $G$ with duplication forest $\Gamma$ and two compatible duplicate sequences $\theta_1$ and $\theta_2$, we have
$\delta(\theta_1)=\delta(\theta_2)$ and
$\enm(\theta_1)=\enm(\theta_2)$.
\end{theorem}

\begin{theorem}
\label{thm:ratio}
Given a network $G$ with duplication history $\Gamma$,  the ratio of two likelihood functions for two duplicate sequences $\theta_1$ and $\theta_2$ that are compatible with $\Gamma$ is given by
$$
\frac{L(\theta_1\,|\,G,\dmc,\Gamma)}{L(\theta_2\,|\,G,\dmc,\Gamma)}
=\Big(\frac{1-p}{2}\Big)^{\lnm(\theta_1)-\lnm(\theta_2)}.
$$
\end{theorem}

\subsection{Reconstruction Algorithms}
\label{sec:alg}

By Theorem~\ref{thm:ratio}, solving Problem 2 is equivalent to solving the following problem.

\begin{problem}
Given a network $G$ and its duplication forest $\Gamma$, construct a duplicate sequence $\theta$ such that
the loss number $ \lnm(\theta)$ is minimized among all sequences compatible with $\Gamma$.
\end{problem}

In this section, we propose some heuristic algorithms to solve Problem 3, and hence Problem 2. The first one is a greedy algorithm called minimal loss number (MLN), {  in which we choose a duplicate node with the  smallest value $\lnm(v)$ among all candidate ones.}

To motivate our main reconstruction algorithm, we introduce some further notation and results.
A duplicate sequence $\theta_1=(v_{1},\cdots,v_n)$ is said to be swapped from $\theta_2=(v'_{1},\cdots,v'_n)$
at position $m$
for some index  $m\in \{1,\cdots,n-1\}$ if we have $v'_m=v_{m+1}$, $v'_{m+1}=v_m$, and $v'_i=v_i$ for all other indices $i$.

 \begin{lemma}
 \label{lem:swap}
Given a network $G$ with duplication forest $\Gamma$, if $\theta_1$ and $\theta_2$ are two compatible duplicate sequences such that $\theta_1$ is swapped from $\theta_2$ at position $m$, then we have
$G^{\theta_1}_i=G^{\theta_2}_i$ for each index $i\in\{0,\cdots,n\}$ with $i\not =m$.
 \end{lemma}

Let $\theta_1$ and $\theta_2$ be two compatible duplicate sequences as stated in the above lemma.
 By Lemma~\ref{lem:swap} and Theorem~\ref{thm:ratio},
 $L(\theta_1\,|\,G,\Gamma,\dmc)\geq L(\theta_2\,|\,G,\Gamma,\dmc)$ if and only if for $G_m=G_m^{\theta_1}=G_m^{\theta_2}$, we have
\begin{equation}\label{ineq:deltacomp}
\lnm(v_m,G_m)+\lnm(v_{m-1},\ro_{v_m}(G_m))\leq \lnm(v_{m-1},G_m)+\lnm(v_m,\ro_{v_{m-1}}(G_m)).
\end{equation}
Motivated by the above observation, for two cherries $\{u,v\}$ and $\{u',v'\}$ in $\Gamma_t$, we say $\{u,v\}$ is more {\em favorable} than $\{u',v'\}$, denoted by $\{u,v\}\fav \{u',v'\}$, if
$
\lnm(v,G_t)+\lnm(v',\ro^{u}_{v}(G_t))< \lnm(v',G_t)+\lnm(v,\ro^{u'}_{v'}(G_t))
$
holds. Note that in general the relation $\fav$ is not transitive, that is,  $\{u,v\}\fav \{u',v'\}$ and  $\{u',v'\}\fav \{u^*,v^*\}$ does not imply  $\{u,v\}\fav \{u^*,v^*\}$.

Now we present our main inference algorithm called cherry greedy (CG), which runs as follows: At every backward reconstruction step, we choose a node from the most favorable cherry $C$, that is, the number of cherries $C'$ with $C\fav C'$ is maximized. If several cherries are equally favorable, we uniformly choose one of them.
 More precisely,  starting from $G_t:=G$ and $\Gamma_t:=\Gamma$,  we choose a most favorable cherry $(u,v)$ from $\Gamma_t$ and uniformly choose one node from the cherry, say $v_t$, as the duplicate node at this step. Then we construct $G_{t-1}$ as $\ro_{v_t}(G_t)$ and $\Gamma_{t-1}=\ro_{v_t}(\Gamma_t)$. This process continues until $G_{0}$ is obtained.

 Since the above algorithm is a stochastic one, that is, among a chosen cherry $\{u,v\}$, $u$ and $v$ has the equal probability of being chosen as the duplicate node. Therefore, one natural way of improving its accuracy is to repeat the algorithm for a certain times and report the best output, {  where the number of repetitions can be regarded as a tuning parameter.} When the real duplicate sequence $\theta_{\real}$ is known, the best one is defined as the output $\theta$ such that Kendall's $\tau$ between $\theta_{\real}$ and $\theta$ is maximized (see Section 3 for further details on Kendall's $\tau$), otherwise  the one with the smallest loss number is chosen. This strengthened version of the CG algorithm with be refereed to as CGR, where `R' stands for repetition.


\subsection{Estimating Parameters}\label{s:estpar}
From the results in Section~\ref{sec:theory} and Section~\ref{sec:alg}, it is clear that the parameters of the DMC model are not used in our inference framework. Moreover, here we will present a method by which
 the parameters can be established after a growth history being inferred.

To this end, assume that a growth history $\hist=(G_0,\cdots,G_n)$, together with the duplicate sequence $(v_{1},\cdots,v_n)$ and anchor list $(u_{1},\cdots,u_n)$, is given. Note that for each neighbor $w$ of node $u_i$ in $G_{i-1}$, the probability that $w$ is adjacent to both $u_i$ and $v_i$ in $G_i$ is $p$. In other words, the extension number $\enm(v_i)$ at $i$-th step, i.e., the number of the common neighbors shared by $u_i$ and $v_i$ in $G_i$, has the binomial distribution with parameters $p$ and $\lnm(u_i)+\enm(v_i)$, where $\lnm(u_i)+\enm(v_i)$ is the number of neighbors that $u_i$ has in $G_{i-1}$.  On the other hand, the random variable $\delta(v_i)$ has Bernoulli distribution with parameter $p_c$. Therefore, we are led to propose the estimators
$
\hat{p}=\frac{\enm(\theta)}{\lnm(\theta)+\enm(\theta)}
$
and 
$
\hat{p}_c=\frac{\delta(\theta)}{n}
$
 to estimate the parameters $p$ and $p_c$ respectively.

\section{Results}\label{sec:result}

Our reconstructing algorithms, minimal loss number (MLN) and cherry greedy (CG), have been implemented in Perl, which is available upon request. Given a network $G$ and duplication forest $\Gamma$, each outputs a hypothetical duplicate sequence $\theta$ that approximates the one with the minimal loss number. 

To assess the performance, we need to measure the difference between the inferred duplicate sequence and the `real' one.
One popular index for this purpose is Kendall's tau $K_\tau$~\cite{bar06,Navlakha11}. Formally, for two sequences
$\theta_1=\{v_{1},\cdots,v_n\}$ and $\theta_2=\{v'_{1},\cdots,v'_n\}$ that consist of the same set of nodes, $K_\tau(\theta_1,\theta_2)$ is defined as
$$
K_\tau(\theta_1,\theta_2)=\frac{2(n_c-n_d)}{n(n-1)},
$$
where $n_c$ is the number of concordant pairs, that is, the number of pairs in $\theta_1$ that are in the correct relative order with respect to $\theta_2$,and $n_d$ is the number of discordant pairs. Note that we have $K_{\tau}(\theta_1,\theta_2)=1$ if the two sequences are identical, and $K_{\tau}(\theta_1,\theta_2)=-1$ if they are exactly opposite.

\subsection{Simulation Validation}
\label{sub:sim}

To validate our algorithms, we generated 100 random network using each DMC model $\dmc$, where the parameters $p_c$ and $p$ ranged from 0.1 to 0.9 at 0.2 intervals.
Each network has 100 nodes and is evolved from the same seed graph $K_2$ (i.e., the graph with two nodes and one edge).

For each simulated network $G$, its duplication forest $\Gamma$ and duplicate sequence $\theta_{\real}$ were recorded. Next, we reconstructed duplicate sequences using our algorithms. The one using MLN is denoted by $\theta_{\mb}$, and the one using CG by $\theta_{\cg}$.
 We also {  considered} the algorithm CGR, which outputs $\theta_{\cgr}$,
 the one with the highest Kendall's $\tau$ among ten runs of CG. 
 {  
 We ran some of the experiments more than 10 times  but
 found that more runs did not improve the results much, and hence 
 we ran 10 times throughout. 
 }
 For comparison, we also generated a random duplicate sequence $\theta_{\rand}$, which can be interpreted as a `null model'. Finally, we computed $K_{\tau}(\theta_{\real},\theta)$ for $\theta\in \{\theta_{\rand},\theta_{\mb},\theta_{\cg},\theta_{\cgr}\}$.

\begin{figure}[h]
    \begin{center}
    \begin{minipage}{0.45\linewidth}
        {\resizebox{\columnwidth}{!}{\includegraphics[scale=1]
        {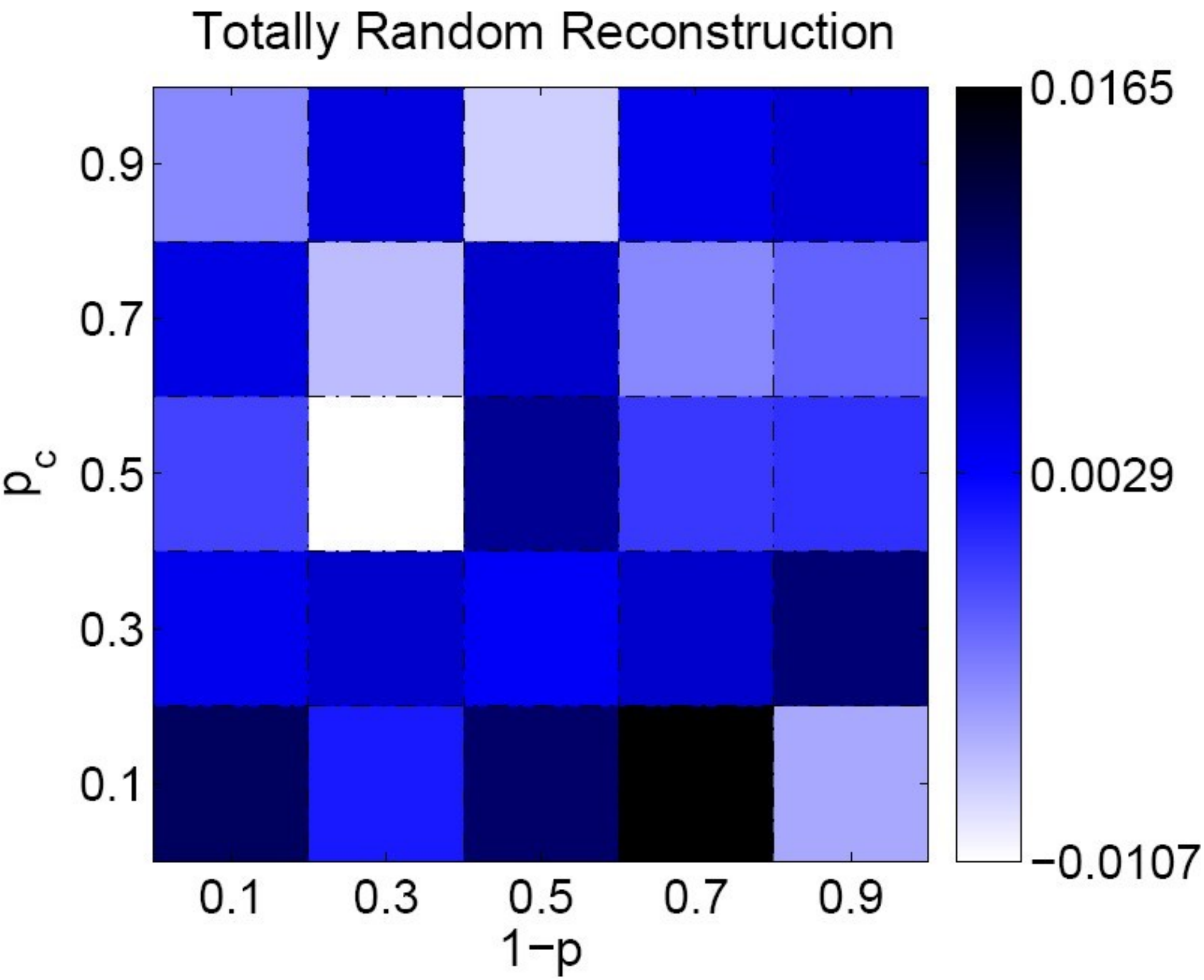}
        }}
    \end{minipage}
    \begin{minipage}{0.45\linewidth}
        {\resizebox{\columnwidth}{!}{\includegraphics[scale=1]{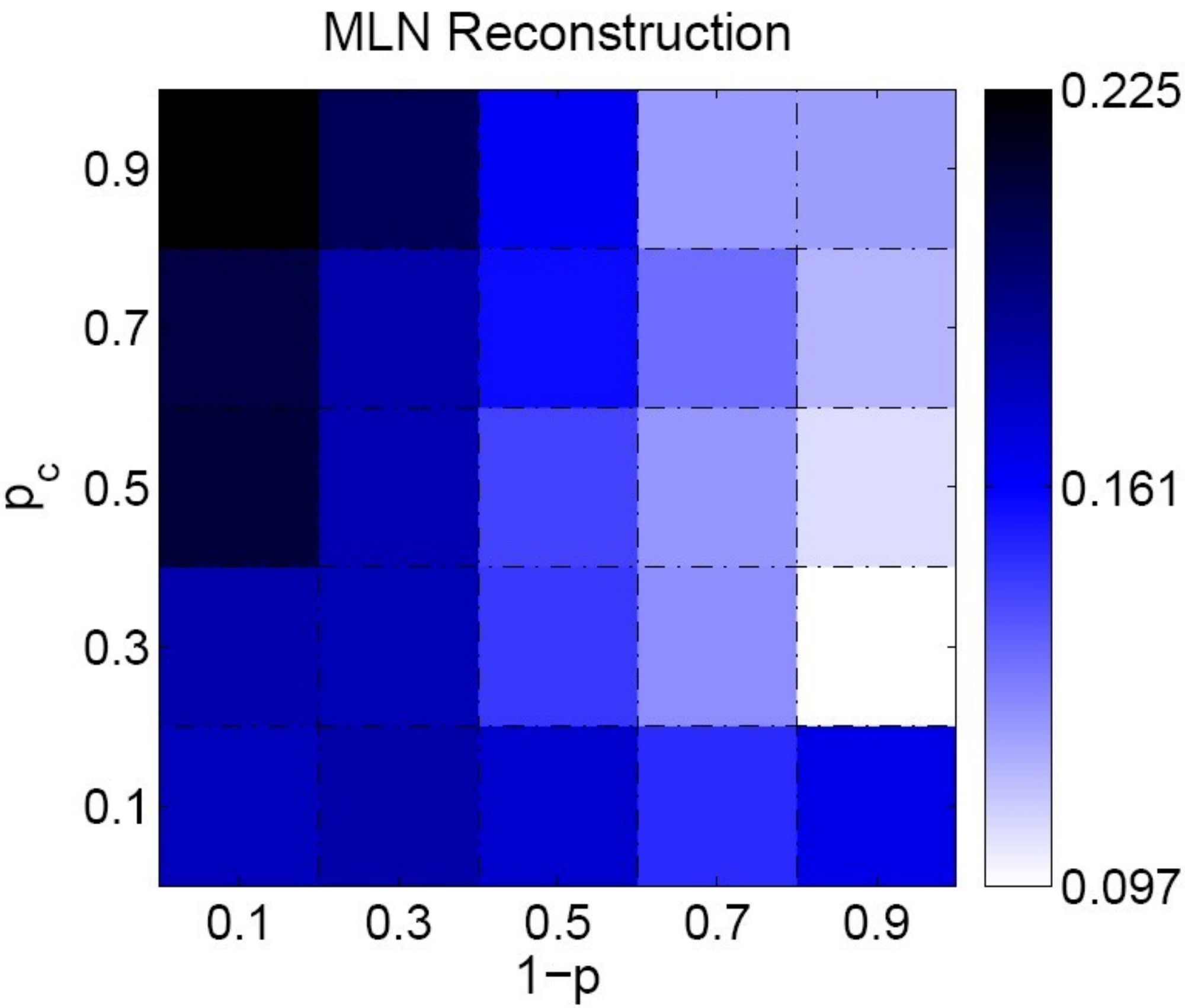}}}
    \end{minipage}
    \caption{
    Results for simulation data sets. The figure in the left is the heat map representing the values of $K_{\tau}(\theta_{\real},\theta_{\rand})$, and the one in the right is for $K_{\tau}(\theta_{\real},\theta_{\mb})$.
    Here the value of Kendall's $\tau$ is represented by the intensity of color.
    }
    \label{fig:Rand:MB}
    \end{center}
\end{figure}

\begin{figure}[h]
    \begin{center}
    \begin{minipage}{0.45\linewidth}
        {\resizebox{\columnwidth}{!}{\includegraphics[scale=1]{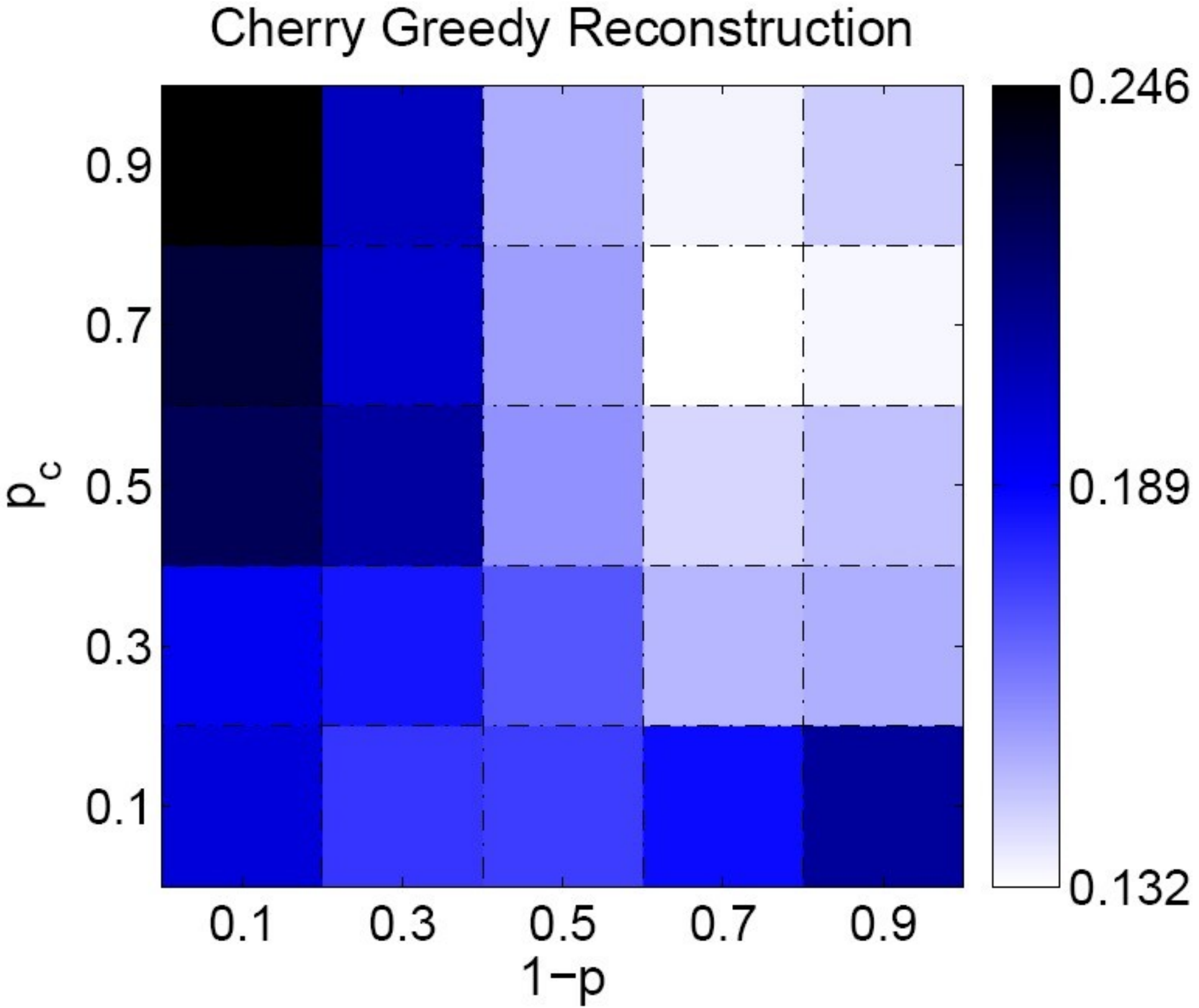}}}
    \end{minipage}
    \begin{minipage}{0.45\linewidth}
        {\resizebox{\columnwidth}{!}{\includegraphics[scale=1]{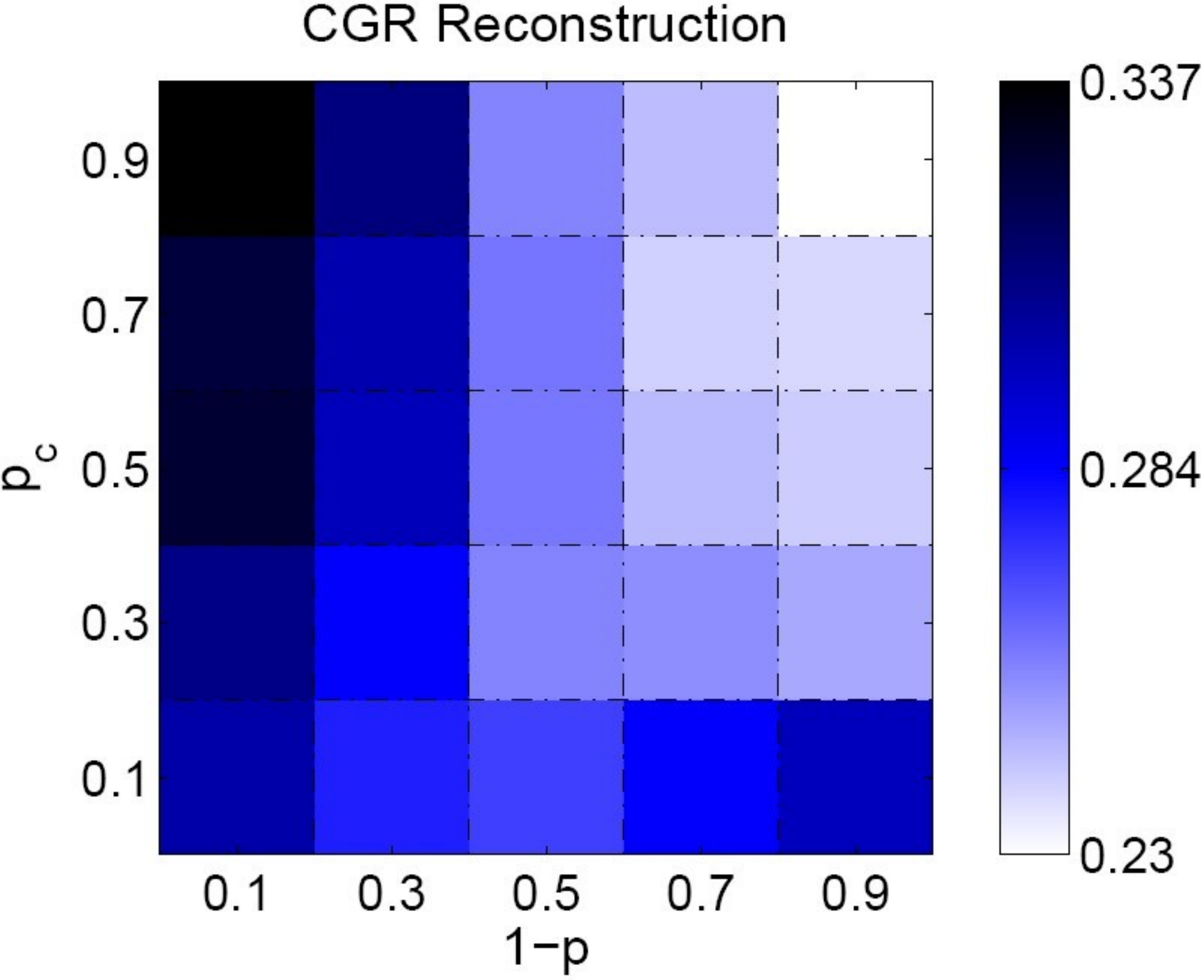}}}
    \end{minipage}
    \caption{Results for the algorithm CG and CGR. The figure in the left is the heat map for $K_{\tau}(\theta_{\real},\theta_{\cg})$ and the one in the right for $K_{\tau}(\theta_{\real},\theta_{\cgr})$. In CGR we run CG for 10 times and report the output with the highest Kendall's $\tau$.}
    \label{fig:CG}
    \end{center}
    \end{figure}

The results for $K_{\tau}(\theta_{\real},\theta_{\rand})$ and $K_{\tau}(\theta_{\real},\theta_{\mb})$ are summarized in Fig.~\ref{fig:Rand:MB}.
Our results for $K_{\tau}(\theta_{\real},\theta_{\rand})$ agree well with the theoretical mean of $K_{\tau}(\theta_{\real},\theta_{\rand})$, which is 0.
In addition, the results for $K_{\tau}(\theta_{\real},\theta_{\cg})$ and $K_{\tau}(\theta_{\real},\theta_{\cgr})$ are summarized in Fig.~\ref{fig:CG}. From these results, we can see that compared to random duplicate sequences, our algorithms have improved the values of Kendall's $\tau$ substantially. In addition, in general CG  has better performance than MLN. Finally, repeating algorithm CG a few times will increase the performance.

\subsection{Comparison with Existing Methods}

In this subsection, we compare the performance of our algorithm CG with NetArch, the inference method introduced in~\cite{Navlakha11}. Since duplication forest is not incorporated in the framework proposed in~\cite{Navlakha11}, it would be expected that CG will outperform NetArch.

Indeed, Fig.~\ref{fig:CG} already shows that our algorithm CGR outperforms NetArch because in~\cite{Navlakha11}, the authors claimed that the values of Kendall's $\tau$ between the real duplicate sequence and the one constructed by their method are between $0.2$ and $0$ for the same set of combinations of parameters.

Even without using repetition, CG also outperforms NetArch in general. We demonstrate this by comparing the performance of them over 100 simulated random networks. For each simulation, we generated a pair of parameters $p$ and $p_c$ uniformly from the interval $(0,1)$, and one graph $G$ with 30 nodes from the seed graph $K_2$ using the DMC model $\dmc$. As above, the duplication forest $\Gamma$ and duplicate sequence $\theta_{\real}$ were recorded. Next,  both  NetArch and CG were used to reconstruct the duplicate sequence, and their outputs were denoted by $\theta_{\rm Net}$ and $\theta_{\cg}$  history.
 Finally, the values $\tau_1:=K_{\tau}(\theta_{\real},\theta_{\cg})$  and $\tau_2:=K_{\tau}(\theta_{\real},\theta_{\rm Net})$ were computed.

\begin{figure}[h]
    \begin{center}
    \subfloat[]{
    \begin{minipage}{0.45\linewidth}
        {\resizebox{\columnwidth}{!}{\includegraphics[scale=1]
        {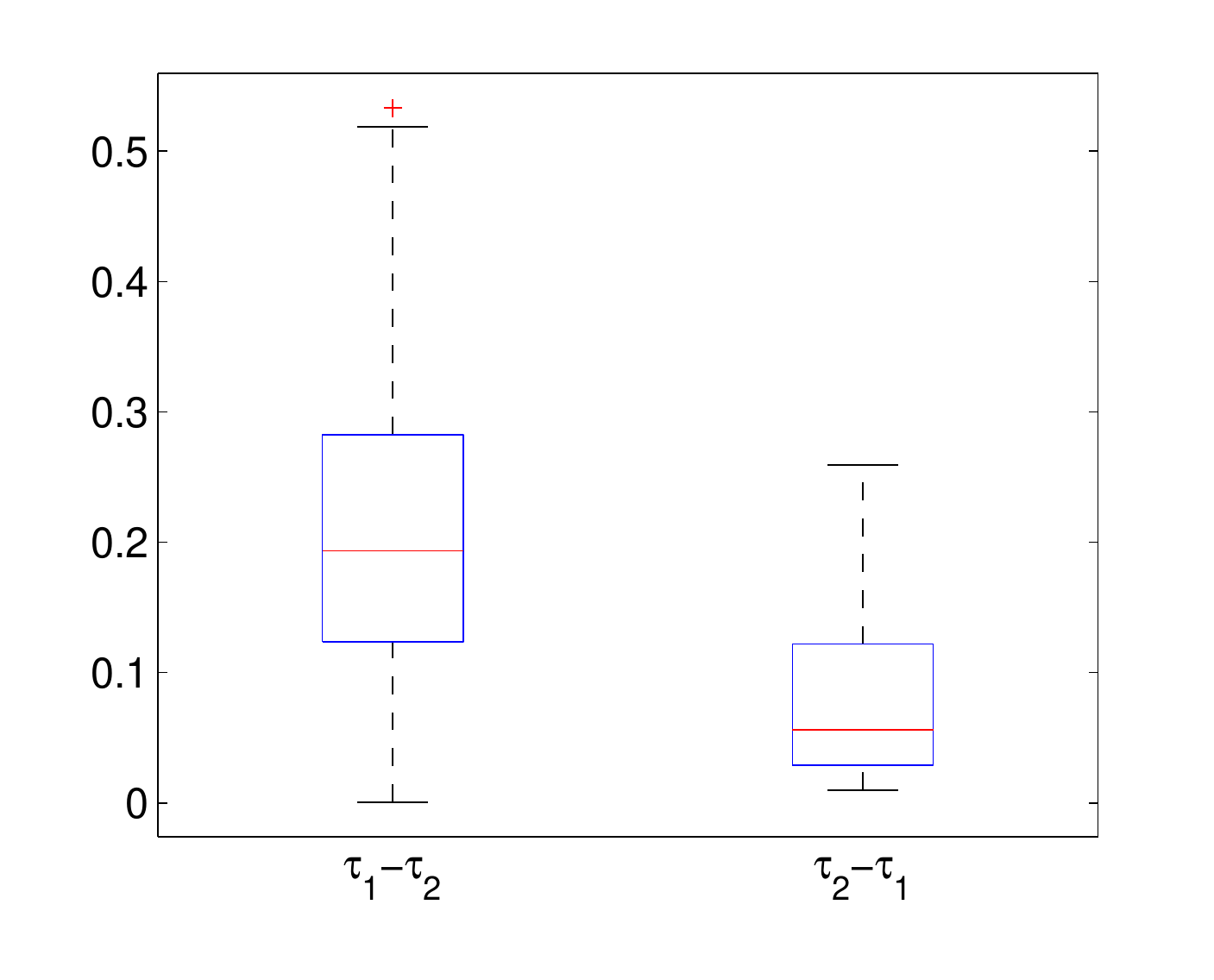}}}
        \label{fig:Kendall_a}
    \end{minipage}
    }
    \subfloat[]{
    \begin{minipage}{0.45\linewidth}
        {\resizebox{\columnwidth}{!}{\includegraphics[scale=1]
        {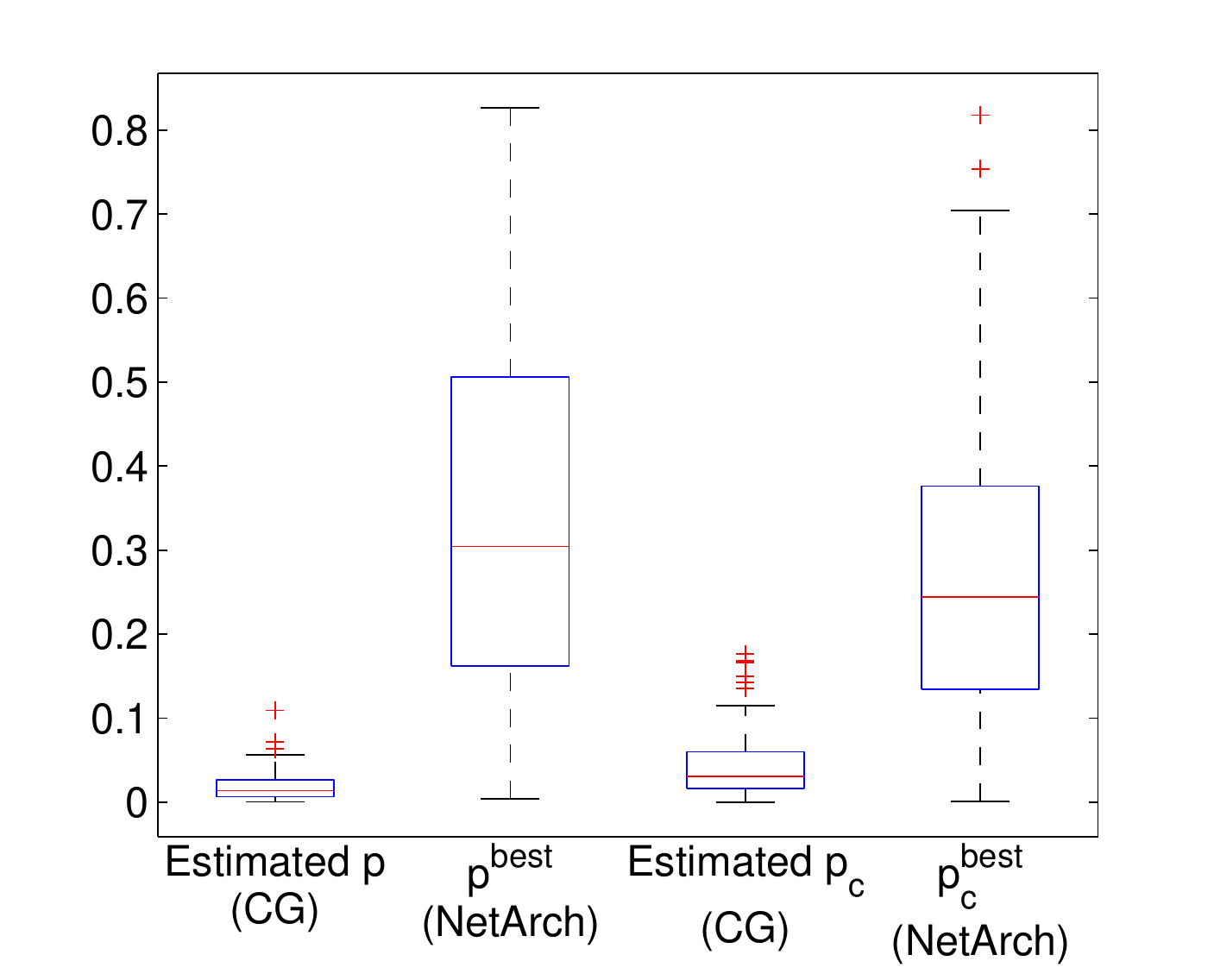}}}
        \label{fig:Kendall_b}
    \end{minipage}
    }
    \caption{(a) Box plot for differences between two methods.  $\tau_1$ is the Kendall $\tau$ obtained by CG and $\tau_2$ is obtained by NetArch. For $\tau_1-\tau_2$, we only consider the cases $\tau_1>\tau_2$, and likewise for $\tau_2-\tau_1$.
     (b) Box plot for errors of estimations of parameters. Here parameters are uniformly generated from the interval $(0,1)$.   }
\end{center}
\end{figure}

Among the 100 simulated networks, CG outperforms NetArch $87$ times, and the distributions of $\tau_2-\tau_1$ and $\tau_1-\tau_2$ are summarized in Fig.~\ref{fig:Kendall_a}.
Note that for the cases when CG outperforms NetArch, the gains in terms of Kendall's tau is significant, i.e., the average value is 0.2.

Moreover, we also compared the parameters $\hat{p}$ and $\hat{p_c}$ estimated by using CG with the ones $p^{best}$ and $p_c^{best}$ obtained by the method in
~\cite{Navlakha11}.
Fig.\ref{fig:Kendall_b} are the box plots for the errors of these four estimations, in which the data are calculated as $|p-\hat{p}|$, etc. Note that the closer to $0$, the better the estimation is. We can see that our method has smaller means of errors and smaller length of confidence intervals for both $p$ and $p_c$.

\subsection{Application to Real PPI Networks}

We downloaded $460$ gene trees reconciled in \cite{dutko07}. The gene trees contain genes from {\em S. cerevisiae} (budding yeast), {\em D. melanogaster} (fruit fly) and {\em C. elegans} (worm).
For each gene tree, we used the genes of one species and deleted all the genes from the other two species to create a gene duplication forest for each species.
In addition, we downloaded corresponding PPI networks from the database DIP
 ( http://dip.doe-mbi.ucla.edu/dip/Main.cgi).
 Since the gene trees obtained in this way are timed, we can infer from them a duplicate sequence ${\theta}^*_{\real}$ that approximates the real duplicate sequence.

 When we checked the gene trees, we found that some of them, especially the large ones, are very asymmetric about the root, which are not common for the duplication trees associated to networks generated by the DMC model. To handle this asymmetry, we modified our inference algorithm CG by taking account the depth of leaves (i.e., the number of edges between the leave and the root). More precisely, in each backward step we choose the most favorable cherry among the cherries whose depth is larger than a threshold. The output of this modified CG algorithm will be denoted by $\theta^*_{\cg}$.

The values of $\tau=K_{\tau}({\theta}^*_{\real}, \theta^*_{\cg})$ for the three networks are listed in Table~\ref{table:app}. In addition, the corresponding estimated parameters $\hat{p}$ and $\hat{p}_c$ are also listed. Note that these estimations are consistent with those in \cite{wagner01,nadia06}, where the authors asserted that $p$ and $p_c$ are smaller than $0.1$.
 Since the one obtained in~\cite{Navlakha11} is $0.7$, here we also demonstrate the advantage of incorporating duplication history in growth history reconstruction.


\begin{table}
\centering
 \caption{The Kendall's $\tau$ and estimated parameters for three PPI networks.}
    \begin{tabular}{|c|c|c|c|}
        \hline
           & {\em S.cerevisiae~~}  & {\em C. elegans~~} & {\em D. melanogaster~~}   \\
        \hline
         $\hat{p}$ & $0.061142$ & $0.020976$ & $0.025953$ \\
        \hline
         $\hat{p}_c$ & $0.053215$  & $0.048443$ & $0.024182$ \\
        \hline
         $\tau$ & $ 0.378 $ & $0.316 $ & $0.473$ \\
        \hline
    \end{tabular}
   \label{table:app}
\end{table}

 \subsection{An improved measure}

 Typically one cannot distinguish between a duplicate node from its anchor node. Therefore, while Kendall's tau between two sequences is natural for comparing duplicate sequence, it also inherits the intricate difficulty of separating anchor nodes from duplicate nodes. To overcome this problem, we propose an alternative measure to compare two duplicate sequences, by which the `symmetry' between anchor nodes and duplicate nodes is taken into account.

\begin{figure}[h]
       \begin{center}
       \begin{minipage}{0.45\linewidth}
           {\resizebox{\columnwidth}{!}{\includegraphics[scale=1]{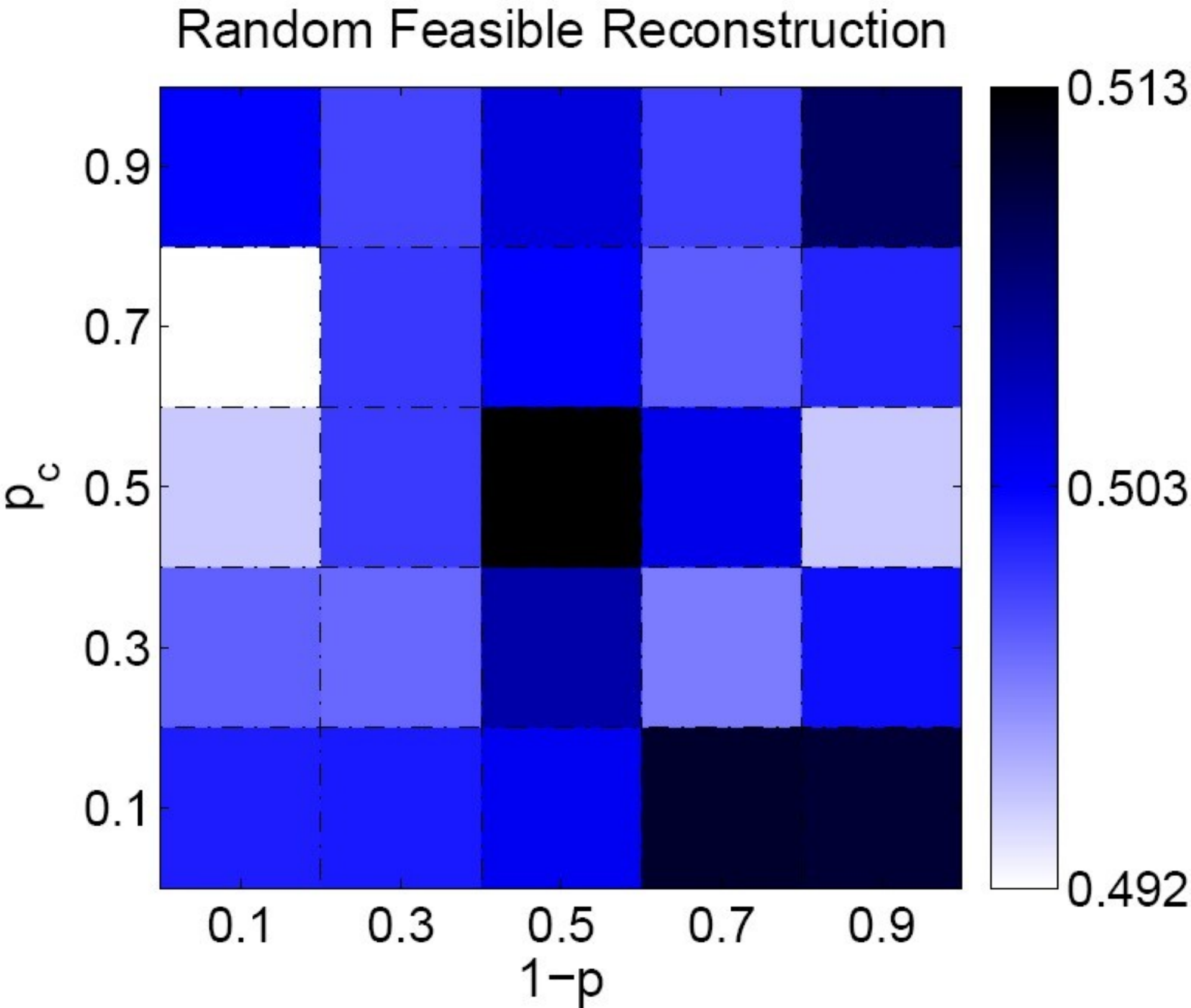}}}
       \end{minipage}
       \begin{minipage}{0.45\linewidth}
           {\resizebox{\columnwidth}{!}{\includegraphics[scale=1]{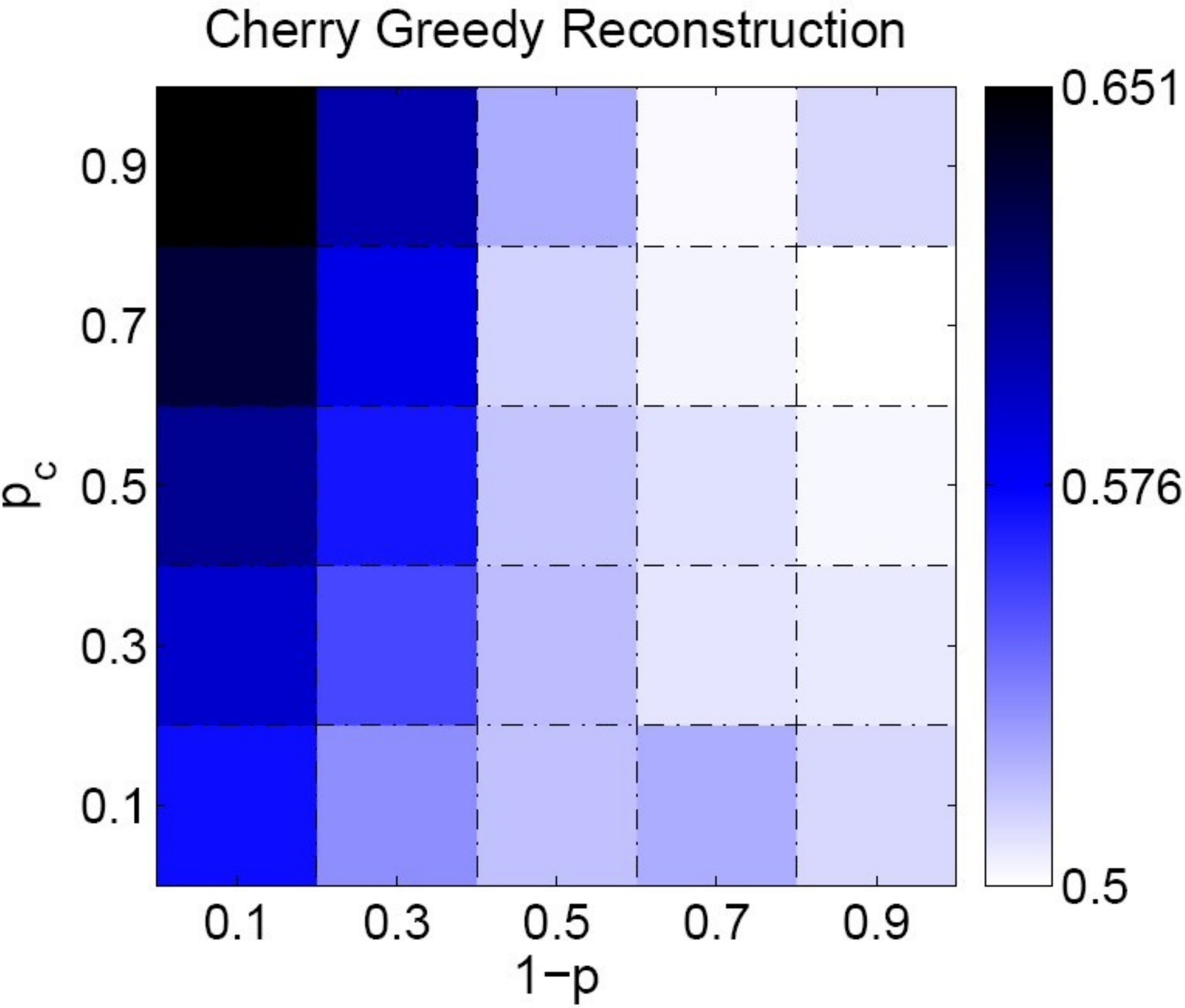}}}
       \end{minipage}
       \caption{     Results measured by $K^*_{\tau}$. The figure in the left is  for $K^*_{\tau}(\theta_{\real},\theta_{{\rm comp}})$ and the one in the right for $K^*_{\tau}(\theta_{\real},\theta_{{\rm \cg}})$. Here the simulated networks are the same as the ones used in obtaining Fig.~\ref{fig:Rand:MB}.     }
       \label{fig:relabel}
       \end{center}
    \end{figure}

    To begin with, each internal node of the duplication forest $\Gamma$ is labeled by a unique label. Note that each duplicate sequence $\theta$ that is compatible with $\Gamma$ induces a unique sequence $\tseq(\theta)$  by replacing duplicate node $v_i$ with the label of the parent of $v_i$ in $\Gamma^{\theta}_i$. For two duplicate sequences $\theta_1$ and $\theta_2$, let $K^*_{\tau}(\theta_1,\theta_2):=K_{\tau}(\gamma(\theta_1),\gamma(\theta_2))$,  and we argue this is a more appropriate measure since here we do not make a distinction between anchor nodes and duplicate nodes. Using   the simulated networks obtained in Section~\ref{sub:sim}, we present in Fig.~\ref{fig:relabel} the results for $K^*_{\tau}(\theta_{\real},\theta_{{\rm comp}})$ and $K^*_{\tau}(\theta_{\real},\theta_{{\rm \cg}})$, where $\theta_{{\rm comp}}$ is a duplicate sequence uniformly chosen from all compatible sequences. These results also validate our algorithm CG as  $K^*_{\tau}(\theta_{\real},\theta_{{\rm \cg}})$ is higher than $K^*_{\tau}(\theta_{\real},\theta_{{\rm comp}})$.

\section{Discussion}\label{sec:discuss}

Assuming the observed network is the result of a growing mechanism as depicted in the DMC model, we have presented a likelihood-based algorithm for recovering the most probable network evolutionary history by exploiting the known duplication history trees of paralogs in the observed network.  Through a series of reduction of the search space of all histories to (i) compatible duplicate sequences and (ii) the set of favored duplicate nodes, we have provided a computationally efficient algorithm. Our approach successfully re-traces the network evolution especially in the scenario that the labels of ancestor nodes are not necessarily to be one of the duplicates. As a useful by-product of our reconstruction, we propose natural estimators for the model parameters which are of independent interest. Our approach can be applied to infer the order of duplication events and to trace the topological characteristics of networks as they evolve. 
 Our method, though described in the context of the DMC model, can be adapted to other network growing models. In addition, it can potentially be extended to predict the emergence of interactions and modules during the network evolution, and hence to provide comparison of the evolution history across different species.

\subsubsection*{Acknowledgments}
This work is supported from the Singapore MOE grant R-146-000-134-112. We are grateful to Dr. Navlakha and Kingsford for providing the code in~\cite{Navlakha11}.



\section*{Appendix}

\noindent
{\em Proof of Lemma~\ref{lem:seed:graph}:}
Assume that $\Gamma$ consists of $k$ binary trees $T_1,\cdots,T_k$, and $\theta$ is a duplicate sequence compatible with $\Gamma$. For each graph $G$ in the graph sequence $\{G_{0}^{\theta},\cdots,G_n^{\theta}\}$, we can associate it with a graph ${\rm \Pi}(G)$ as follows. The vertex set of ${\rm \Pi}(G)$ is $\{1,\cdots,k\}$ and two distinct vertices $i$ and $j$ are adjacent if and only if there exist some adjacent nodes $g_i$ and $g_j$ in $G$ such that $g_i$ is a leaf in the tree $T_i$ and $g_j$ is a leaf in $T_j$.

Let $G$ be a graph in $\{G_{1}^{\theta},\cdots,G_n^{\theta}\}$. Denote the anchor node and duplicate node corresponding to this graph by $u$ and $v$, respectively. Since $\theta$ is compatible, $u$ and $v$ are the leaves in the same tree in $\Gamma$. Note that for any vertex $g$ that is distinct from $u$ and $v$, then $g$ is adjacent to $u$ or $v$ in $G$ if and only if $g$ is adjacent to $u$ in $\ro_v^u(G)$.
Therefore, we can conclude that ${\rm \Pi}(G)={\rm \Pi}(\ro_v^u(G))$, and hence also
${\rm \Pi} (G^{\theta}_{0})={\rm \Pi} (G^{\theta}_n)$.  On the other hand, from the construction we know that  ${\rm \Pi}(G^{\theta}_{0})$ is isomorphic to $G^{\theta}_{0}$.

In consequence, for two compatible duplicate sequences $\theta_1$ and $\theta_2$, since $G^{\theta_1}_{n}=G_n=G^{\theta_2}_{n}$, we can conclude that
$G^{\theta_1}_{0}$ and $G^{\theta_2}_{0}$ are isomorphic, as required.
\hfill $\square$

\bigskip
\noindent
{\em Proof of Theorem~\ref{thm:comp:num}:}
We shall establish the lemma by induction on the number of cherries in $\Gamma$. The base case that $\Gamma$ contains no cherry is trivial, because this implies $n=0$.

Now assume that $\Gamma$ contains $m$ cherries, and that the lemma holds when the number of cherries in the duplication forest is at most $m-1$. Fix a cherry $\{u,v\}$ in $\Gamma$ and choose a label $g$ that is not used before. Consider the network $G^*$ that is obtained from $\ro_v^u(G)$ by relabeling $u$ with $g$, and the duplication forest ${\Gamma}^*$ obtained from $\Gamma$ by replacing the cherry $\{u,v\}$ with a leaf labeled as $g$. Note that either node $u$ or $v$ (possible both) must appear in the duplicate sequence of $\theta_1$; we replace them with $g$ and denote the sequence with the first $g$ removed by $\theta_1^*$. Then $\theta_1^*$ is a duplicate sequence that is compatible with ${\Gamma}^*$.

 Similarly, the sequence ${\theta_2^*}$ obtained from $\theta_2$ in the same way is also compatible with ${\Gamma}^*$.
Now the induction assumption implies
$ \delta(\theta_1^*)= \delta({\theta_2^*}).
$
Together with
$$
\delta({\theta_1})-\delta({\theta_1^*})
=\delta({\theta_2})-\delta({\theta_2^*}),
$$
we have $ \delta({\theta_1})= \delta({\theta_2})$, as required.

On the other hand, the number of edges increased from $G^{\theta}_{i-1}$ to $G^{\theta}_i$ is given by $\delta(v_i)$ and $\enm(v_i)$, where $v_i$ is the duplicate node. Together with Lemma~\ref{lem:seed:graph}, this implies
$$\delta(\theta_1)+\enm(\theta_2)
=|E(G_n)|-|E(G^{\theta_1}_{0})|
=|E(G_n)|-|E(G^{\theta_2}_{0})|
=\delta(\theta_2)+\enm(\theta_2).$$
Since $\delta({\theta_1})= \delta({\theta_2})$, we have
$ \enm({\theta_1})= \enm({\theta_2})$.
\hfill $\square$

\bigskip
\noindent
{\em Proof of Theorem~\ref{thm:ratio}:}
Let $\theta=\{v_{1},\cdots,v_n\}$ be a duplicate sequence that is compatible with the duplication forest $\Gamma$.
By Lemma~\ref{lem:seed:graph} and Theorem~\ref{thm:comp:num}, it is sufficient to note that
$$
L(\theta\,|\,G,\dmc,\Gamma)=p_c^{\delta(\theta)} p^{\enm(\theta)} q^{\lnm(\theta)},
$$
holds with $q:=(1-p)/2$, an observation following from that
$$
\pr(G_i^{\theta}\,|G_{i-1}^{\theta},\Gamma,\dmc)
=p_c^{\delta(v_i)}p^{\enm(v_i)}q^{\lnm(v_i)}
$$
holds for each $i\in \{1,\cdots,n\}$.
\hfill $\square$

\bigskip
\noindent
{\em Proof of Lemma~\ref{lem:swap}:}
Clearly, we have $G_i^{\theta_1}=G_i^{\theta_2}$ for $i>m$. To show this also holds for $i<m$, it suffices to show  $G_{m-1}^{\theta_1}=G_{m-1}^{\theta_2}$
 For $i\in \{m,m+1\}$, let $u_i$ be the anchor node of $v_i$. Since $\theta_1$ and $\theta_2$ are both compatible with $\Gamma$,  we know that $\{u_m,v_m\}$ and $\{u_{m+1},v_{m+1}\}$ are two distinct cherries in $\Gamma^{\theta_1}_{m+1}=\Gamma^{\theta_2}_{m+1}$. Therefore, we have
$$
\ro^{u_m}_{v_m}(\ro^{u_{m+1}}_{v_{m+1}}(G_{m+1}))
=\ro^{u_{m+1}}_{v_{m+1}}(\ro^{u_{m}}_{v_{m}}(G_{m+1})),
$$
because the four nodes $u_m$, $v_m$, $u_{m+1}$ and $v_{m+1}$ are distinct.
\hfill $\square$

\end{document}